\documentclass[runningheads]{llncs}

\usepackage[utf8]{inputenc}
\usepackage[T1]{fontenc}
\usepackage{textcomp}
\usepackage{graphicx}
\usepackage{subcaption} 
\usepackage{float}%
\usepackage{xcolor}
\usepackage{color}
\usepackage{xspace,tcolorbox}
\usepackage{tikz}
\usetikzlibrary{shapes,arrows,patterns,positioning} 
\usepackage{caption}
\usepackage[normalem]{ulem}
\usepackage{hyperref} 
\usepackage{tabularx}

\usepackage{empheq}
\usepackage[mathscr]{euscript}
\DeclareSymbolFont{rsfs}{U}{rsfs}{m}{n}
\DeclareSymbolFontAlphabet{\mathscrsfs}{rsfs}
\usepackage{mathtools}
\usepackage{stmaryrd} 

\usepackage{dutchcal} 

\usepackage{amsmath, amssymb, amsfonts}
\usepackage{mathtools}
\usepackage{mathrsfs}
\usepackage{bm}

\usepackage{pgfplots}
\pgfplotsset{compat=newest}

\usepackage[utf8]{inputenc}
\usepackage[T1]{fontenc}
\usepackage{textcomp,mathcomp}
\usepackage{xcolor,colortbl} 

\usepackage{chngcntr} 

\newcommand{\halfcheckmark}{{$\checkmark$}\textsuperscript{\textcolor{black}{\kern-0.53em{\bf--}}}}

\usepackage{threeparttable}

\usepackage{booktabs}
\usepackage{nicematrix}

\usepackage{multirow}
\usepackage[ruled,linesnumbered]{algorithm2e}

\usepackage{breakcites}
\usepackage{enumitem}

\usepackage{xargs} 
\usepackage[colorinlistoftodos,prependcaption,textsize=tiny]{todonotes}
\newcommandx{\commentt}[2][1=]{\todo[linecolor=red,backgroundcolor=red!25,bordercolor=red,#1]{#2}}

\makeatletter
\newlength\min@xx
\newcommand*\xxrightarrow[1]{\begingroup
  \settowidth\min@xx{$\m@th\scriptstyle#1$}
  \@xxrightarrow}
\newcommand*\@xxrightarrow[2][]{
  \sbox8{$\m@th\scriptstyle#1$}  
  \ifdim\wd8>\min@xx \min@xx=\wd8 \fi
  \sbox8{$\m@th\scriptstyle#2$} 
  \ifdim\wd8>\min@xx \min@xx=\wd8 \fi
  \xrightarrow[{\mathmakebox[\min@xx]{\scriptstyle#1}}]
    {\mathmakebox[\min@xx]{\scriptstyle#2}}
  \endgroup}
\makeatother

\makeatletter
\newcommand*\xxleftarrow[1]{\begingroup
  \settowidth\min@xx{$\m@th\scriptstyle#1$}
  \@xxleftarrow}
\newcommand*\@xxleftarrow[2][]{
  \sbox8{$\m@th\scriptstyle#1$}  
  \ifdim\wd8>\min@xx \min@xx=\wd8 \fi
  \sbox8{$\m@th\scriptstyle#2$} 
  \ifdim\wd8>\min@xx \min@xx=\wd8 \fi
  \xleftarrow[{\mathmakebox[\min@xx]{\scriptstyle#1}}]
    {\mathmakebox[\min@xx]{\scriptstyle#2}}
  \endgroup}
\makeatother

\usepackage[left=4cm, right = 4cm, top=4cm, bottom=4cm, asymmetric]{geometry}

\begin{document}

\title{MFAz: Historical Access Based Multi-Factor Authorization}

\author{Eyasu Getahun Chekole \and
Howard Halim \and
Jianying Zhou}
\authorrunning{E. G. Chekole et al.}
%
\institute{Singapore University of Technology and Design, Singapore\\
\email{\{eyasu\_chekole,howard\_halim,jianying\_zhou\}@sutd.edu.sg}}

\maketitle

\begin{abstract}

Unauthorized access to private resources (such as data and services) remains one of the most critical security challenges in the realm of 
cybersecurity. With the increasing sophistication of attack techniques, the threat of unauthorized access is no longer confined to the conventional ones, such as exploiting weak access control policies. Instead, advanced exploitation strategies, such as session hijacking-based attacks, are becoming increasingly prevalent, posing serious security concerns. Session hijacking enables attackers to take over an already established session between legitimate peers, thereby gaining unauthorized access to private resources. These attacks are typically conducted in a stealthy manner, making their detection exceedingly difficult. Unfortunately, traditional access control mechanisms, such as static access control policies, 
are insufficient to prevent session hijacking or other advanced exploitation techniques. In this work, we propose a new multi-factor authorization (MFAz) scheme that proactively mitigates unauthorized access attempts 
both conventional and advanced unauthorized access attacks. The proposed scheme employs fine-grained access control rules (ARs) and verification points (VPs) that are systematically generated from historically granted accesses as the first and second authorization factors, respectively. 
As a proof-of-concept, we implement the scheme using different techniques. We leverage bloom filter to achieve runtime and storage efficiency (even for resource-constrained devices), and blockchain to make authorization decisions in a temper-proof and decentralized manner. 
To the best of our knowledge, this is the first formal introduction of a multi-factor authorization scheme, which is orthogonal to and distinct from the widely used multi-factor authentication (MFA) schemes. The effectiveness of our proposed scheme is experimentally evaluated using 
a smart city testbed involving different devices with varying computational capacities. The experimental results reveal high effectiveness 
of the scheme both in security and performance guarantees. 

\end{abstract}

\keywords{Access Control, Multi-Factor Authorization, Multi-Factor Security, Session Hijacking, Blockchain, Decentralized Security}

\section{Introduction}\label{introduction}

In today's digital era, the reliance on digital services has 
increased exponentially. However, the 
rapid proliferation of cyber attacks targeting such services has 
has become a growing concern among stakeholders. In particular, unauthorized access attacks \cite{sloan2017unauthorized,mfaa}, where 
adversaries seek to gain illicit access to protected resources, 
constitute one of the major security concerns in the realm of cybersecurity. 

Unauthorized access attacks can generally be classified into two primary categories: conventional and advanced. In the former case, adversaries exploit vulnerabilities in traditional access control policies, models, and mechanisms to gain unauthorized access. In the latter case, adversaries employ advanced exploitation techniques that allow them to gain unauthorized access. 
Session hijacking attacks \cite{ogundele2020detection,manjula2021pre} represent a prominent example of the latter, wherein adversaries compromise active sessions established between communicating parties (e.g., server and client) with the intent of capturing the session token or session identifier (SID), thereby obtaining unauthorized access to protected resources. An SID is a unique token assigned by a server to a client (user) upon the initiation of a session. It acts as a ticket or credential permitting the user to access authorized resources for the duration of the session. Adversaries may employ a range of techniques to obtain the SID, including session fixation \cite{session_fixation}, cross-site scripting (XSS) \cite{xss}, 
brute forcing \cite{brute_forcing}, and session sidejacking \cite{sauber2017session}. If successful, the adversary could impersonate legitimate users and take over their active session, thereby gaining unauthorized access to restricted resources.

A wide range of access control and authorization solutions have been proposed over the years to combat unauthorized access attacks. The widely adopted solutions are conventional access control models \cite{sandhu1998role,hu2015attribute,zong2019policy,armando2014selective}, which enforce different access control policies to safeguard systems or resources against unauthorized accesses. However, while these solutions could be effective against conventional unauthorized access attacks, they can be bypassed by advanced exploitation strategies, such as session hijacking. 
 Since session hijacking occurs after a session has been established between the parties and access has been granted to legitimate users in accordance with access control policies, conventional access control measures are inadequate in preventing such attacks.

Another widely adopted method to prevent unauthorized access is the implementation of multi-factor authentication (MFA) schemes \cite{banyal2013multi, ometov2018multi, wee2024excavating,angcsur2025}. Although MFA schemes are principally designed to serve authentication functions, some scholars argue that they also contribute to addressing authorization concerns by impeding impersonation attacks, thereby limiting adversaries' ability to obtain unauthorized access. Nevertheless, this approach remains ineffective against session hijacking attacks, as such attacks occur subsequent to the completion of the authentication process.

Various countermeasures have also been developed against session hijacking-related attacks. For example, the industrial standard TLS 1.3 protocol \cite{dowling2021cryptographic} offers certain features to mitigate the likelihood of session hijacking. However, TLS 1.3 does not provide a mitigation strategy once a session is hijacked. In fact, TLS 1.3 mainly addresses network-layer vulnerabilities \cite{singh2022handshake} and does not mitigate application-layer vulnerabilities, such as cross-site scripting (XSS), which can be exploited to hijack SIDs \cite{muzammil2024unveiling}. 

Other session hijacking countermeasures are generally based on time-out and re-authentication mechanisms \cite{ogundele2020detection}, pre- and post-authorization requests \cite{manjula2021pre}, multi-layer data encryption \cite{kaiser2022multi}, and authentication of transposition-encrypted user information (ATEUI) \cite{hwang2022web}. Nonetheless, these approaches remain insufficient to defend against session hijacking-based unauthorized access attacks (see details in Section \ref{related_work}). Moreover, existing authorization approaches do not offer proactive mitigation strategies against session hijacking attacks. Furthermore, none of them introduce a multi-factor authorization strategy to strengthen the defense against unauthorized access attacks. Therefore, it is imperative to propose advanced authorization techniques that can effectively address such security risks.  

In this work, we propose a new multi-factor authorization (MFAz) scheme to alleviate both conventional and advanced unauthorized access concerns discussed above. The proposed scheme employs fine-grained access control rules (ARs) and verification points (VPs) as first and second authorization factors, respectively. The ARs are based on attribute-based fine-grained access control rules \cite{armando2014attribute} and are primarily used to restrict conventional unauthorized access attacks. On the other hand, the VPs are systematically generated from historically granted accesses 
and nonces (e.g., timestamps) and are used as a second layer of defense to proactively mitigate session hijacking-based unauthorized accesses. Access is granted only when both ARs and VPs are successfully verified. 

To make our scheme applicable in different settings with varying device capacities, we have also paid careful attention to efficiency, availability, and decentralization when designing and implementing it. To this end, we use bloom filter (BF) and blockchain in the implementation of our scheme. The BF allows us to efficiently store and verify the generated VPs, achieving high storage and runtime efficiency, thus making the scheme practical even for resource-constrained devices. The use of blockchain allows us to immutably store the ARs and VPs in a decentralized manner across different devices in a distributed system. Furthermore, it allows us to achieve greater transparency among the users/devices, user anonymity, and high availability (by avoiding a single point of failure). Finally, we tested the scheme using an IIoT-based 
smart-city testbed that involves different devices with varying computational capabilities, such as edge gateways, PCs, and Raspberry PIs.  

Overall, the proposed scheme addresses not only conventional access control issues but also advanced unauthorized access threats highlighted above. In general, this work makes the following key contributions: 
\begin{itemize}
    \item We propose a multi-factor authorization (MFAz) scheme to address both conventional and session hijacking-based attacks. To achieve this, we employ fine-grained ARs and VPs as first and second authorization factors, respectively. To the best of our knowledge, this is the first work to introduce a multi-factor authorization scheme (which is orthogonal to MFA schemes) against a wide range of unauthorized access attacks.
    \item Our scheme proactively mitigates session hijacking attacks regardless of the technique used to hijack the session. 
    \item The adoption of BF allows us to achieve both runtime and storage efficiency, making our scheme practical even for resource-constrained devices.
    \item The adoption of blockchain allows us to immutably store the ARs and VPs and enables us to make authorization decisions in a decentralized and transparent manner in distributed systems.  
    \item We evaluate the effectiveness of our proposed scheme (covering both its security and efficiency guarantees) using a realistic smart-city based testbed equipped with different devices of varying capabilities. 
\end{itemize}

\section{Related work}\label{related_work}

\subsection{Access control models}

A wide range of conventional access control models and mechanisms, such as role-based  \cite{sandhu1998role}, attribute-based \cite{hu2015attribute,armando2014attribute}, and policy-based \cite{zong2019policy}, have been adopted to safeguard systems and resources against various unauthorized accesses. While these models addressed several types of access control issues in the past, they also suffered various shortcomings. For example, the design of weak access control rules and misconfiguration of policies resulted in catastrophic cyberattacks in various occasions \cite{almushiti2023investigation}. Moreover, the static and inflexible nature of these mechanisms make them ineffective against advanced authorization attacks, such as session hijacking-based unauthorized accesses.     

\subsection{Resilience against session hijacking attacks}

Ogundele et al. \cite{ogundele2020detection} proposed a session time-out and re-authentication for authorized sessions to prevent session hijacking attacks. This method provides a layer of protection against a user who leaves authenticated sessions unattended. However, suppose a malicious user is actively targeting this session and able to get the session cookies and hijack the session before time-out. In that case, it does not have any other security mechanism that can be relied on.

Manjula et al. \cite{manjula2021pre} proposed a methodology to prevent session hijacking attacks using pre-authorization and post-authorization requests on a web server. The pre-authorization process validates the session between client and server through a session-ID. The post-authorization process distinguishes between two types of requests: web API requests, which are read-only, and session requests, which are read and write. This approach enhances security by detecting malicious activities that attempt to exploit authorized session requests. However, it has limited flexibility, as it is specifically designed for read-write web applications and may not be effective against session hijacking in other contexts.

Sathiyaseelan et al. \cite{sathiyaseelan2017proposed} proposed a scheme to prevent session hijacking using modified one-time cookies (OTCs) and a reverse proxy server (RPS). The RPS acts as an intermediary between the client and the server, filtering and forwarding all client and server requests. The RPS verifies each request using OTC, IP address, session ID, and browser fingerprinting. It generates a new OTC and disposes of the old one each time a request passes through the RPS for verification. However, the paper does not evaluate the workload required to dispose of and regenerate the OTC with every access request. Additionally, the scheme has limited usage because it relies on IP addresses and browser fingerprints to generate the OTC. As a result, other authorization processes that do not involve a browser cannot utilize this scheme.

Hwang et al. \cite{hwang2022web} proposed a defense mechanism called authentication of transposition-encrypted user information (ATEUI) to prevent session hijacking using user-specific information. This user information includes multiple identifiers such as BIOS serial numbers, device login account information, mainboard serial numbers, UUID values of devices, and a random number. This methodology is capable of detecting malicious access requests even when the same IP and MAC addresses of an authorized user are used, due to the unique user information employed. However, due to the rigorous mechanism, it demonstrates significantly slower performance. According to the results in \cite{hwang2022web}, compared to other common techniques, ATEUI takes approximately ten times longer to complete the process.

\subsection{TLS 1.3 and secure cookies}

The discussion of session hijacking attacks can be broadly categorized into two layers of the OSI model -- the network layer and the application layer\cite{kamal2016state}. As an SSL/TLS protocol, TLS 1.3 \cite{dowling2021cryptographic} offers several features to mitigate the likelihood of session hijacking. However, like its predecessors, TLS 1.3 does not provide protection once a session is compromised (e.g., if a session ID is leaked due to poor cookie implementation). While TLS 1.3 effectively addresses network-layer vulnerabilities, such as packet sniffing, downgrade attacks, and weak cryptographic algorithms \cite{singh2022handshake}, it does not mitigate application-layer vulnerabilities, such as cross-site scripting (XSS), which can be exploited to hijack SIDs \cite{muzammil2024unveiling}. Hence, TLS 1.3 does not provide the necessary prevention for our threat models.

Therefore, in addition to implementing TLS 1.3, web services also adopt secure cookies to address 
the limitations of TLS. HTTP cookies \cite{kristol2001http} have become a widely adopted component in web client-server communication, serving as a cornerstone for handling authentication and authorization mechanisms \cite{cahn2016empirical}. Cookies store essential information for web browsing, such as SIDs, which are used to authenticate and authorize clients. Cookie hijacking has become a common threat leading to data leaks, primarily due to the use of unencrypted and insecure cookie implementations. Several studies \cite{squarcina2023cookie,sivakorn2016cracked,dacosta2012one,kwon2019security} have demonstrated that securing cookies with the \texttt{HttpOnly} flag, which makes the cookie inaccessible to JavaScript on the client side, and \texttt{Secure} flag, which ensures the cookie is only transmitted over encrypted connections (HTTPS), can significantly enhance security and help mitigate the risk of session ID leakage from cookies. 

Therefore, by combining TLS 1.3 and secure cookies, session hijacking attacks could be 
minimized due to encrypted and secure communication between the web client and server. However, 
our threat model is not only focus on
preventing the initial establishment of session hijacking. 
But it also focuses on how to proactively mitigate the impact once the session has already been compromised in anyway (i.e., the attacker has already acquired the session ID). 
TLS 1.3 and secure cookies do not provide such proactive mitigation strategies against session hijacking attacks.

\subsection{Multi-factor authorization}

To the best of our knowledge, there are no proper multi-factor authorization (MFAz) schemes (apart from MFA schemes) designed in prior works. However, there are certain multi-layer related schemes available in the literature. For example, Alsahlani et al. \cite{alsahlani2021lmaas} proposed a lightweight multi-factor authentication and authorization model (LMAAS-IoT), which only utilizes a one-way hash function and bitwise XOR operation. Fuzzy extractor algorithm is used to assist in verifying user-side biometric information. 
However, this is mainly an MFA scheme, not an MFAz scheme. It does not address concerns about session hijacking-based attacks.

Addobea et al. \cite{addobea2023secure} introduced a novel approach to mitigate single points of failure inherent in most designs using a cloud-centric model. Their proposed solution is an access control system that leverages the blockchain technology to authenticate users by integrating multiple factors into a time-based access code for generating user private keys. This model addresses the limitations of non-distributed architectures, such as those based on cloud models. However, even if it is an access control system, the multi-factor aspect is for authentication, not for authorization. This scheme also 
does not provide any mechanism to detect or prevent malicious activities, including session hijacking attacks.

Kaiser et al. \cite{kaiser2022multi} proposed a fully secured data flow system using IAM (Identity Access Management), MFA (Multi-Factor Authentication), IDAAS (Identity as a service), AAAS (Authorization as a service), SAML (Security Assertion Markup Language), and data encryption using AES. This design relies on the combination of its heavily layered pre-existing security mechanism. Although it enhances security through heavy layering of the mechanisms, it is not a proper MFAz scheme. It also does not address 
session hijacking-based unauthorized access attacks.

Yao et al. \cite{yao2020dynamic} proposed a Zero-Trust architecture for dynamic access control and authorization, utilizing a trust-based access control (TBAC) model. The model employs trust and authorization nodes to assign roles and permissions to users. Trust calculation is derived from the variance between historical and current user behavior. This variance is then compared against predefined trust thresholds set by the system to determine the user authorization's status. However, the user portrait employed in the architecture is susceptible to subjectivity, which can lead to undesired outcomes. 

\section{Preliminaries}\label{preliminaries}
\subsection{Notation}

All relevant notations used throughout this paper are summarized in Table \ref{tab:notation}. 
\begin{table}[htb]
    \centering
    \vspace{-0.8cm}
    \caption{Description of notations}
     \label{tab:notation}
    \begin{tabularx}{\linewidth}{lX}
        \hline
        \textbf{Symbol} & \textbf{Description} \\ \hline
        $AU$ & Authorized user \\ 
        $UU$ & Unauthorized user \\
        $GA_{i}$ & Historical granted accesses of user $i$ \\
        $GA_{i}^{j}$ & The $j^{th}$ session of $GA_{i}$ \\ 
        $SGA_{i}$ & A randomly selected GAs from $GA_{i}$ \\
        $VP_{i}$ & Verification point of user $i$\\
        $VP_{i}^{j}$ & The $j^{th}$ session of $VP_{i}$ \\
        $K_{i}$ & A long-term authentication key of $U_{i}$\\
        $U_{i}$ & User $i$ \\
        $O_{i}$ & Operation $i$ \\
        $R_{i}$ & Resource $i$ \\
        $T_{s}$ & Timestamp \\
        $\mathcal{H}(\cdot)$ & A cryptographic hash function \\
        \hline
    \end{tabularx}
     \vspace{-1cm}
\end{table}

\subsection{Bloom filter}

Bloom filter (BF) is a space-efficient probabilistic data structure used to represent sets of elements \cite{luo2018optimizing}. Introduced by Burton Bloom \cite{bloom1970space}, the BF is designed to address the space/time trade-off inherent in hash coding methods by allowing a controlled rate of errors (false positives). The efficiency of BF in performing fast membership queries is further elaborated in \cite{broder2004network}. BF algorithm primarily involves three operations: initialization, insertion, and verification.

\subsection{Blockchain}
Blockchain is a distributed and immutable ledger that records transactions using a peer-to-peer network connection \cite{khan2021blockchain}. Satoshi Nakamoto introduced bitcoin \cite{nakamoto2008bitcoin} as the first cryptocurrency distributed system. Following bitcoin, many new distributed cryptocurrency systems were introduced, such as Ethereum. Ethereum introduced smart contracts into its blockchain system, enabling users to deploy executable programs within blockchain transactions \cite{buterin2014next}.

Smart contracts are automated executable protocols that verify, validate, and enforce the terms (contract) made by two or more parties within the blockchain environment \cite{wang2019blockchain}. Moreover, smart contracts deployed in the blockchain environment are transparent and immutable, ensuring tamper-proof execution. 

\section{Threat model}\label{threat_model}

Our threat model focuses on both conventional access control threats and advanced unauthorized access techniques. In the former, we focus on attacks that rely on weak access control policies or mechanisms. In the latter, we focus on advanced attack techniques (e.g., session hijacking attacks) that attempt to gain unauthorized access to protected resources. 

\subsection{Conventional access control threats} \label{access_control_attacks}

In this section, we discuss conventional access control threats that mainly target weak access control policies or methods. Access control policies are essential for managing system security, as they regulate each request made by a client, determining whether to approve or deny the request \cite{samarati2000access}. Over the years, various access control models have been developed, including role-based access control (RBAC), mandatory access control (MAC), discretionary access control (DAC), zero-trust architecture, and identity-based access control, among others. 
However, they all face a common issue: broken access control (BAC).

BAC is a vulnerability that allows an unauthorized user ($UU$) to access private resources 
due to the weaknesses in the established access control policy. This vulnerability can occur for several reasons, such as misconfigured platforms, unprotected functionality, static files, insecure access control methods, identifier-based functions, and multistage functions \cite{almushiti2023investigation}. 
An adversary can exploit these and other vulnerabilities to gain unauthorized access to some or all of the protected resources in the system. A high-level architecture of such attacks is illustrated in Figure \ref{fig_weak_access_control}.

\subsection{Session hijacking threats}\label{session_hijacking}

\begin{figure}[htb]
     \centering
     \begin{subfigure}{0.40\textwidth}
         \centering
         \includegraphics[width=\linewidth]{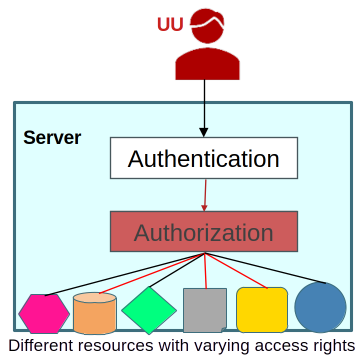}
         \caption{}
         \label{fig_weak_access_control}
     \end{subfigure}
     \begin{subfigure}{0.50\textwidth}
         \centering
         \includegraphics[width=\linewidth]{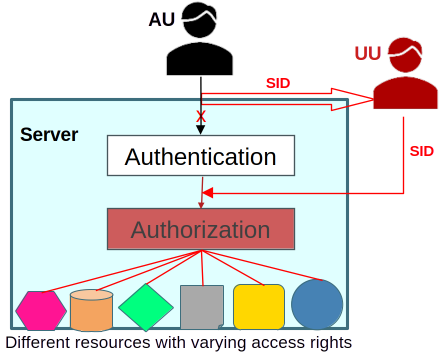}
         \caption{}
         \label{fig_session_hijacking}
     \end{subfigure}
     \caption{Threat model: (a) weak access control threats (b) session hijacking threats}
     \label{fig_threat_model}
\end{figure}

In this section, we discuss session hijacking threats, where an $UU$ systematically takes over an active session established between two legitimate users, e.g., an authorized user ($AU$) and a server, to gain unauthorized access to protected resources. A high-level architecture of this threat is illustrated in Figure \ref{fig_session_hijacking}. 
In this threat, the $UU$ aims to obtain the 
session ID (SID) of the parties via either a client-side (mostly) or server-side (rarely) web applications or other services. 
This can be achieved by employing various techniques, including session fixation, cross-site scripting (XSS), brute forcing, and session sidejacking \cite{sauber2017session}. 
\begin{itemize}
    \item \textbf{Session fixation} \cite{session_fixation} is a technique where the $UU$ compels the $AU$ to authenticate using a predefined malicious SID chosen by the $UU$. As a result, the $AU$ unknowingly enters an authenticated session using this malicious SID, which is already known to the $UU$, thereby allowing the attacker to gain unauthorized access once the session is established.
    \item \textbf{XSS attack} \cite{xss} is a technique that allows attackers to exploit vulnerabilities within a web application to inject malicious scripts into webpages viewed by $AU$. In this technique, the $UU$ injects malicious JavaScript code into a webpage that is rendered by the $AU$'s browser. Once the malicious script is executed, it can steal the SID stored in the $AU$'s cookies, URL or elsewhere. 
    \item \textbf{Brute forcing} \cite{brute_forcing} is a technique where $UU$ attempts to predict or guess an active SID through trial and error or by analyzing certain patterns.
    \item \textbf{Session sidejacking} \cite{sidejacking} is a technique that allows $UU$ to obtain the SID by intercepting and reading the network traffic between $AU$ and the server. This is often possible when the communication channel is unsecured. 
\end{itemize}

Once the SID is acquired using any of the techniques discussed above, $UU$ initiates a replay attack by leveraging the pre-authenticated SID to communicate with the server. This grants $UU$ access to the authenticated session, allowing them to execute malicious actions within the context of the authorized session. $UU$ could also launch a DoS or other attack techniques on $AU$ to end its session with the server, thereby avoiding detection by preventing two simultaneous uses of the same SID. 

\subsection{Threat assumptions}

In general, we assume the following in our threat model.
\begin{itemize}
    \item The generated historical granted accesses (GAs) are securely stored in the $AU$'s device, and $AR$s and $VP$s on the blockchain. We assume that the adversary cannot access the GAs. 
    \item The communication channel between $AU$ and the server is properly secured using SSL/TLS or other security protocols. Hence, the adversary cannot sniff or intercept any information exchanged between the two parties, including the GAs.
    \item In case of the conventional access control threats, the adversary can exploit weak access control rules. 
    \item In case of the session hijacking threats, the adversary can obtain the SID using various techniques, such as session fixation, cross-site scripting (XSS), and brute forcing, but not via session sidejacking as the communication channel is secured.
    \item After obtaining the SID, the adversary may also terminate $AU$'s session with the server (e.g., using DoS or other techniques) and place itself instead. This allows $UU$ to avoid potential 
    suspicion or detection by preventing two simultaneous uses of the same SID.
\end{itemize}

\section{MFAz: Proposed Scheme}\label{proposed_scheme}

\subsection{Overview}
In this section, 
we discuss our proposed multi-factor authorization (MFAz) scheme that is designed to address both conventional and advanced (e.g., session hijacking) unauthorized access attacks. 
The proposed scheme employs access control rules ($AR$) and systematically generated verification points ($VP$) as the first and second authorization factors. 
The former is derived from the conventional 
attribute-based access control model \cite{hu2015attribute}. The latter is systematically generated from historically granted accesses 
and nonces (e.g., timestamps). To achieve efficiency and further security goals, we leverage bloom filter (BF) and blockchain when implementing the scheme.  



MFAz is designed based on a smart-city based testbed (see Section \ref{experimental_setup} for details) that involves different devices, including edge gateways, cloud servers, PCs and RPIs. To simplify our presentation, we use a client-server architecture where we specify \emph{user space} (denoting all users who try to access resources from the servers) and \emph{server space} (denoting all devices 
that act as a server and 
share resources to the users). Motivated by the architecture of the the smart-city testbed we used, 
we use blockchain to securely store $AR$ and $VP$ 
as smart contracts. 
The proposed scheme has two main phases: \emph{user enrollment} and \emph{access authorization}.

\subsection{Enrollment phase} \label{initialization}

User enrollment takes place when a new user is registered with any of the servers. This procedure involves two main processes: credential provisioning and system bootstrapping. In the former, the user submits its cryptographic credential, e.g., its long-term authentication key $K_{i}$, and relevant user attributes to the server. Since the authentication and key verification processes are orthogonal processes to the access authorization task we are focusing on, we omitted the details in that regard. Upon receiving the long-term key $K_{i}$, the server 
initializes the user's entry in the blockchain. During the $VP$ bootstrapping process, the server generates certain dummy $GA$s and $VP$s for the user. 
While the $VP$s are inserted in the BF and then securely stored in the blockchain, 
the $GA$s are securely stored in the user's device. These two processes will facilitate the conditions to effectively enforce the proposed MFAz policy for subsequent access requests of an already enrolled user. 

\subsection{Authorization phase}

\begin{figure*}[htb]
    \centering
     \includegraphics[scale=0.5]{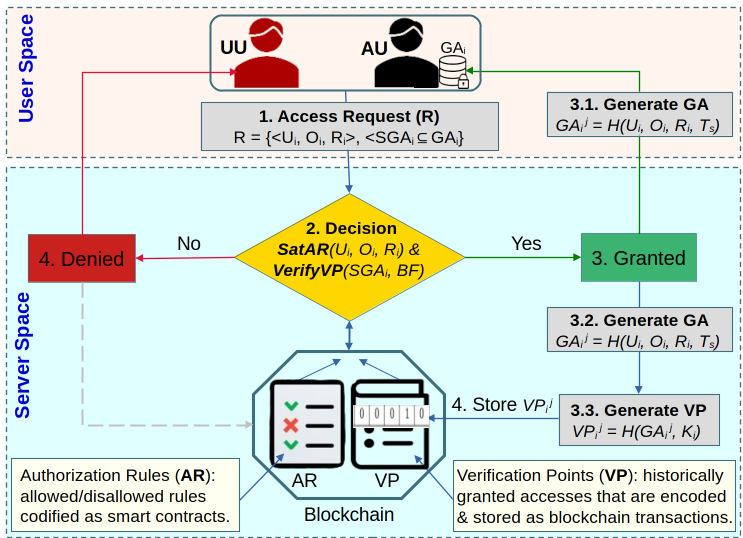}
    \caption{An illustration of the authorization process in MFAz}
    \label{fig_mfaz_architecture}
\end{figure*}

As discussed in the preceding sections, our proposed 
scheme involves $AR$s and $VP$s as authorization factors to raise the bar against unauthorized users from getting access to private resources. The objective of this work is to restrict both conventional access control threats, where attackers exploit weaknesses in access control policies, and session hijacking threats, wherein attackers compromise established sessions to impersonate legitimate users and gain unauthorized access to resources. A high-level workflow of the authorization process is illustrated in Figure \ref{fig_mfaz_architecture} and discussed in detail as follows. 
\begin{enumerate}
    \item First, a user (from the user space) sends access request (R) to a server (in the server space) to access a specific resource. This request consists of the following information: the user attribute $U_i$ (e.g., authentication key $K_i$), the type of operation to be performed $O_i$ (e.g., read, write and execute), the resource to be accessed $R_{i}$, and a few list of historically granted accesses $SGA_{i}$ that are randomly selected from locally stored $GA$s. In conventional systems, a user may present only the session ID (SID) to gain access to the resource, provided the SID remains valid. However, this approach does not differentiate between a legitimate user attempting to re-access the resource after a brief pause and an attacker who may have obtained the SID through illicit means. Our proposed scheme mitigates this risk and prevents unauthorized access, as discussed below.
    \item Upon receiving the access request, the server 
    makes an informed decision. 
    First, the server runs the $\mathsf{SatAR()}$ algorithm (discussed in Section \ref{mfaz_algorithms}) and checks whether the attributes provided by the user satisfies the $AR$s stored in the blockchain as a smart contract. If it is satisfiable, the server performs the next verification. It runs the $\mathsf{VerifyVP()}$ algorithm (cf. Section \ref{mfaz_algorithms} and \textbf{Algorithm} \ref{alg:vp_verification} for details) to check whether the user provided $SGA_{i}$ 
    can be verified. This verification is performed by first converting $SGA_{i}$ into $VP_{i}$ using the $\mathsf{GenVP()}$ algorithm (cf. \textbf{Algorithm} \ref{alg:vp_generation}), followed by checking the presence of $VP_{i}$ in the bloom filter. 
    \item If both 
    checks satisfied, the server grants the requested access to the user. Then, 
    the following tasks will be performed subsequently:
    \begin{enumerate}
        \item The user locally generates $GA_{i}^{j}$ (i.e., the $j^{th}$ $GA$ of user $i$) using the $\mathsf{GenGA()}$ algorithm (refer the definition in Section \ref{mfaz_algorithms}) and securely store it for future use.
        \item The server also generates $GA_{i}^{j}$ similarly. Using $GA_{i}^{j}$ and the 
        user's key $K_i$ as inputs, it then generates the respective $VP$ (i.e., $VP_{i}^{j}$) using the $\mathsf{GenVP()}$ algorithm (cf. Section \ref{mfaz_algorithms}). 
        \item Then, the server inserts the $VP$ in the bloom filter $BF$ using the $\mathsf{InsertVP()}$ algorithm (cf. Section \ref{mfaz_algorithms}) and then stores it on the blockchain using the $\mathsf{StorBF()}$ algorithm (cf. Section \ref{mfaz_algorithms}). 
    \end{enumerate}
    \item If any of the authorization mechanisms fail, the server detects the attempt as malicious activity and 
    it rejects the requested access. 
\end{enumerate}

 

\subsection{Algorithms and operations of MFAz}\label{mfaz_algorithms}

In this section, we describe the main algorithms or operations used in the user enrollment and authorization processes of MFAz. 

\begin{itemize}
     \item $\mathsf{0/1 \gets SatAR(U_i, O_i, R_i)}$: This algorithm takes the user, operation and resource attributes as inputs and returns $1$ if the access control rules specified in $AR$ are satisfied, and 0 otherwise.
     
    \item $\mathsf{GA_{i}^{j} \gets GenGA(U_{i}, R_{i}, O_{i}, T_{s})}$: This algorithm is run by the user and the server to generate $GA_{i}^{j}$ (i.e., the $j^{th}$ $GA$ of user $i$) by taking attributes $U_{i}$ (e.g., authentication key $K_{i}$), operation $O_{i}$, resource $R_{i}$, and the timestamp $T_{s}$ as inputs. The user generated  $GA_{i}^{j}$ is securely stored in its device to be used as a token in for future access requests while the server generated  $GA_{i}^{j}$ is subsequently used to generate the respective  $VP_{i}^{j}$ and get discarded. The pseudocode is provided in \textbf{Algorithm} \ref{alg:ga_generation}
    
    \item $\mathsf{VP_{i}^{j} \gets GenVP(GA_{i}^{j}, K_{i})}$: This algorithm is run by the server to generate $VP_{i}^{j}$ (i.e., the $j^{th}$ $VP$ of user $i$) by taking $GA_{i}^{j}$ and $K_{i}$ as inputs. The pseudocode is provided in \textbf{Algorithm} \ref{alg:vp_generation}.
    
    \item $\mathsf{0/1 \gets CheckVP(VP_{i}, BF)}$: This algorithm is run by the server to check whether a verification point $VP_{i}$ is in the bloom filter $BF$. It returns 1 if $VP_{i}$ is in  $BF$, and 0 otherwise.
    
    \item $\mathsf{0/-1 \gets InsertVP(VP_{i}, BF)}$: This algorithm is run by the server to insert a verification point $VP_{i}$ to the bloom filter $BF$. It returns 0 if $VP_{i}$ is successfully inserted to $BF$ or it is already there, and -1 if there is an error. 
    
    \item $\mathsf{0/1 \gets VerifyVP(VP_{i}, BF)}$: This algorithm is run by the server to check whether a particular $VP$ is in the BF. 

     \item $\mathsf{StoreBF(BF, BC)}$: This algorithm is run by the server to store the $BF$ (or the updated $BF$ after inserting new $VP$s) on the blockchain ($BC$). 
     This is achieved through the invocation of smart contract methods. These smart contracts enable the system to automatically trigger a new event with the updated $BF$ value, adding the event to the transaction ledger. This approach ensures a decentralized and secure authorization process using blockchain.

    \item $\mathsf{FetchBF(BF, BC)}$: This algorithm is run by the server to fetch the latest $BF$ from the $BC$ via a smart contract method. This method is particularly important when we verify $VP$s. Because, we need to first fetch the latest $BF$ from the $BC$ to check the presence of the $VP$ in the $BF$.
    %
\end{itemize}

\begin{minipage}{0.51\linewidth}
\begin{algorithm}[H]
\caption{$VP$ generation algorithm $\mathsf{GenVP()}$}\label{alg:vp_generation}
\SetAlgoLined
\KwIn{$U_{i}$, $O_{i}$, $R_{i}$, 
$T_{s}$, $K_{i}$}
\KwOut{$VP_{i}^{j}$}
Generate $GA_{i}^{j}$ $\gets$ $\mathsf{GenGA}(U_{i}, O_{i}, R_{i}, 
T_{s})$\;
Generate $VP_{i}^{j}$ $\gets$  $\mathcal{H}(GA_{i}^{j},K_{i})$\;
Insert $\mathsf{InsertVP}(VP_{i}^{j}, BF)$\;
Store $BF$ $\rightarrow$ blockchain\;
\Return $VP_{i}^{j}$\;
\end{algorithm}
\end{minipage}
\hfill
\begin{minipage}{0.42\linewidth}
\begin{algorithm}[H]
\caption{$GA$ generation algorithm $\mathsf{GenGA()}$}\label{alg:ga_generation}
\SetAlgoLined
\KwIn{$U_{i}$, $O_{i}$, $R_{i}$, 
$T_{s}$}
\KwOut{$GA_{i}^{j}$}
Generate $GA_{i}^{j}$ $\gets$ $\mathcal{H}(U_{i}, O_{i}, R_{i}, 
T_{s})$\;
\Return $GA_{i}^{j}$\;
\end{algorithm}
\end{minipage}

\begin{algorithm}
\caption{$VP$ verification algorithm $\mathsf{VerifyVP()}$}\label{alg:vp_verification}
\SetAlgoLined
\KwIn{Access Request comprising $U_{i}$,$O_{i}$, $R_{i}$, 
and $SGA_{i}$}
\KwOut{Access Granted, Access Denied}
Server receive Access Request from user $U_{i}$\;
First Authorization $\rightarrow$ Access Control Rules (AR)\;
\If{AR is not satisfied}{
    \Return Access Denied\;
}
\Else{
    Fetch latest $BF$ from Blockchain\;
    Generate $VP_{Temp}$ $\gets$ $\mathsf{GenVP}(SGA_{i}, K_{i})$\;
    \eIf{$\mathsf{CheckVP}(VP_{Temp}, BF)$}{
        $GA_{i}^{j} \gets \mathsf{GenGA}(U_{i}, O_{i}, R_{i}, 
        T_{s})$\;
        $VP_{i}^{j} \gets \mathsf{GenVP}(GA_{i}^{j}, K_{i})$ (see \textbf{Algorithm} \ref{alg:vp_generation})\;
        \Return Access Granted\; 
    }{
        \Return Access Denied\;
    }
}
\end{algorithm}

\subsection{Implementation details}
In this section, we 
discuss the implementation details of MFAz, mainly on the second authorization factor (i.e., $VP$). Firstly, the proposed scheme is built with C/C++ using the MIRACL cryptography library \cite{miracl}, which is essential for the big number implementation and also offers fast and precise crypto algorithms. 
We used SHA-256 as our hash function. 

For runtime and storage efficiency, the proposed scheme uses 
$BF$ to efficiently store and verify the verification points. The $BF$ is initialized and set to accommodate a set number of entries with a set percentage of false positive rate. The entries and false positive rates are subject to future changes depending on usage and the system's performance capabilities. 

For additional security and decentralization gains, we leverage blockchain to implement MFAz. 
It allows us to immutably store $AR$s and $VP$s across different 
devices in our smart-city network. In particular, we utilize ganache \cite{ganache} to provide a local blockchain environment as a testnet for this project.



\section{Evaluation}\label{evaluation}

\subsection{Experimental setup}\label{experimental_setup}
\begin{figure}[htb]
    \centering
     \includegraphics[scale=0.33]{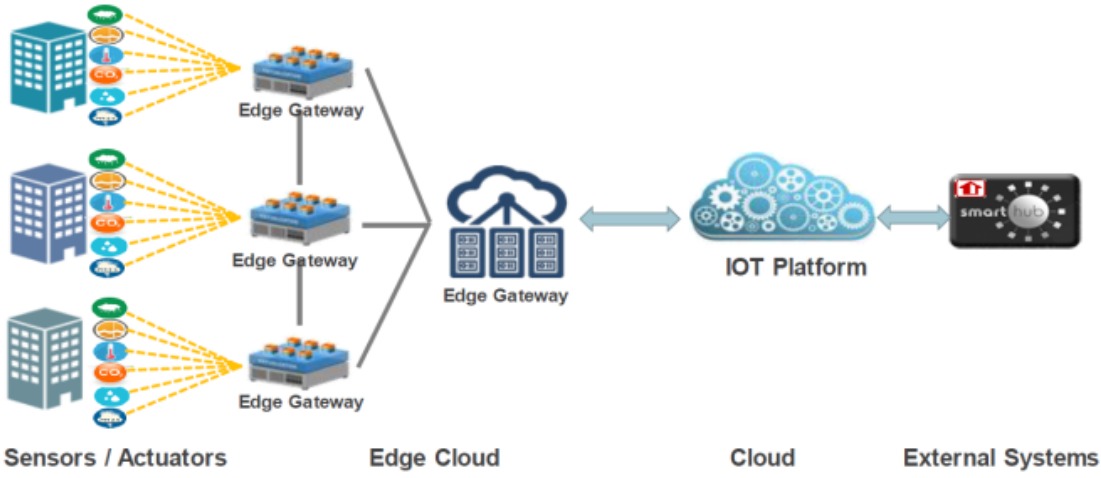}
    \caption{Architecture of the IoT-based smart city testbed}
    \label{fig_estatesense_architecture}
\end{figure}

To stress test the effectiveness of our proposed scheme, we test it using a realistic smart-city based 
testbed (a high-level architecture of the testbed is provided in Figure \ref{fig_estatesense_architecture}). The testbed comprises different devices, including IoT edge gateways, PCs, field devices (sensors and actuators), IoT cloud platform, operators and third-party applications. Different communication protocols, such as HTTP/HTTPS, TCP, UDP, and MQTT  (Message Queuing Telemetry Transport), are also used to exchange messages or transfer files among the devices. To assess the runtime performance of the scheme, we run it on the following devices (acting as servers) representing different hardware and software specifications. 
\begin{itemize}
    \item \textbf{Gateway}:
We use an ADLINK technology MVP-510A embedded fanless computers, a 64-bit system with SMP (symmetric multiprocessing) and vsyscall32 capabilities, as our IoT edge Gateway. It features a motherboard with 16 GB of RAM. The CPU is an intel(R) core(TM) i5-9500TE, operating at a base frequency of 2.20 GHz with a maximum turbo frequency of 3.60 GHz. The software specification for the experimental setup includes ubuntu 22.04 as the operating system. The setup also utilizes node.js version 18.19.1, python version 3.10.12, and g++ version 11.4.0.
\item \textbf{PC}: We use a Dell XPS 13 9315 laptop with a 12th-generation intel core i7 processor and 16GB of RAM. This system operates on Ubuntu 24.04. 
\item \textbf{RPi}: We use Raspberry Pi 4 Model B (with a Quad Core Cortex-A72 of 1.5GHz CPU and 4GB RAM).
\end{itemize}

For this experiment, the $BF$ is initialized to accommodate approximately 1000 entries with a 1\% false positive rate. This configuration results in a $BF$ data structure consisting of approximately 9585 bits and utilizes 7 hash functions. This setup effectively balances memory efficiency with the need to maintain a low false positive rate, ensuring optimal performance for the specified number of entries.

Additionally, a local blockchain is set up using ganache, providing the test account and test network for the ETH blockchain environment. This setup allows for the simulation and testing of blockchain interactions in a controlled environment, ensuring that the system's functionality can be thoroughly evaluated.

\subsection{Security evaluation and discussion}\label{security_evaluation}
In this section, we discuss the achieved security guarantees of the proposed scheme. The evaluation is conducted based on our adversarial model discussed in section \ref{threat_model}.

\subsubsection{Resilience against conventional attacks}

We tested the scheme using valid and invalid user attributes and access control rules. It accurately detects the malicious attempts and it correctly granted access for the legitimate attempts. Therefore, there were no any error observed with the access control rule related attacks we performed.  

\subsubsection{Resilience against session hijacking attacks}

The proposed scheme was rigorously evaluated using various session hijacking techniques. For instance, tested the scheme using both real and fake SIDs, as well as combinations of real and fake $GA$s. The scheme successfully denied access to all malicious attempts while correctly granting access to all legitimate ones. Throughout the evaluation, no errors, either false positives or false negatives, were observed. In the future, we intend to perform more advanced attacks utilizing malware and other sophisticated techniques. 

\subsubsection{Security of VPs Vs SIDs}

As 
discussed, we employed 
$VP$s/$GA$s as authorization factors to mitigate session hijacking and other attacks. 
One may, however, 
argue how $VP$s/$GA$s provide greater resilience compared to 
SIDs. SIDs possess a web-oriented nature, often being embedded within URLs or other web services, which makes them frequent targets for attackers. In fact, numerous techniques already exist to compromise SIDs. In contrast, $VP$s/$GA$s are generated locally and securely stored on a local machine (cf. Figure \ref{fig_mfaz_architecture} about the $GA$s). The only instance in which $GA$s leave the local device is when they are deliberately selected by the user for access authorization. Even then, despite the assumption of a secure communication channel, these $GA$s are destroyed immediately after serving 
once as authorization factors, thus limiting their exposure. Consequently, $VP$s/$GA$s present a reduced attack surface compared to SIDs. Moreover, $VP$s/$GA$s offer enhanced verifiability and suitability as authorization factors because they are intrinsically linked to the user. Their generation is based on the user’s credentials and historical activities, such as the specific resources accessed, operations performed, and corresponding timestamps, further strengthening their integrity and trustworthiness.


\subsection{Performance evaluation}\label{performance_evaluation}

In this section, we evaluate the runtime 
of MFAz (only the $VP$) based on its computation time taken during the $VP$ generation and verification processes using the experimental setup discussed in Section \ref{experimental_setup}. The results are based on an average of 50 runs of the authorization process.

\begin{figure*}[htb]
      \centering    
      \vspace{-0.4cm}
     \begin{subfigure}{0.328\textwidth}
         \includegraphics[width=\linewidth]{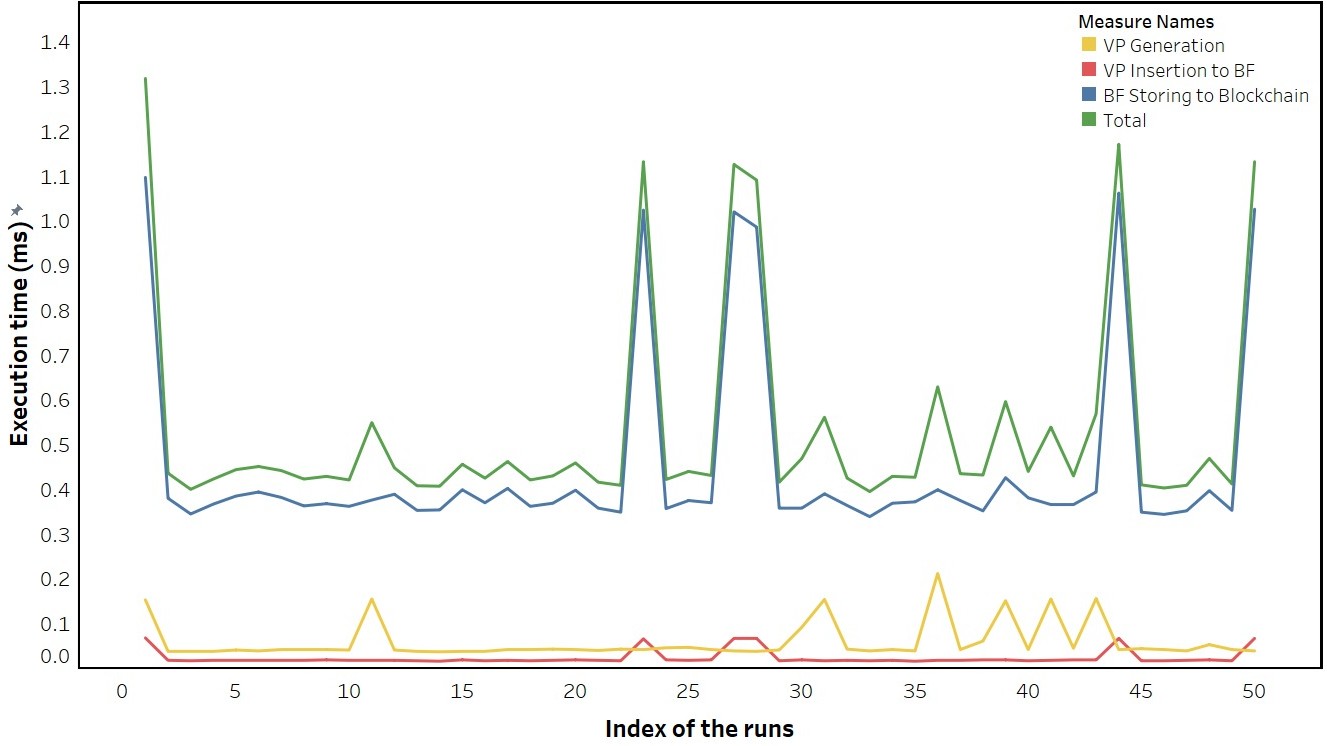}
         \caption{}
         \label{fig_vp_generation_gateway}
     \end{subfigure}
     \begin{subfigure}{0.328\textwidth}
         \includegraphics[width=\linewidth]{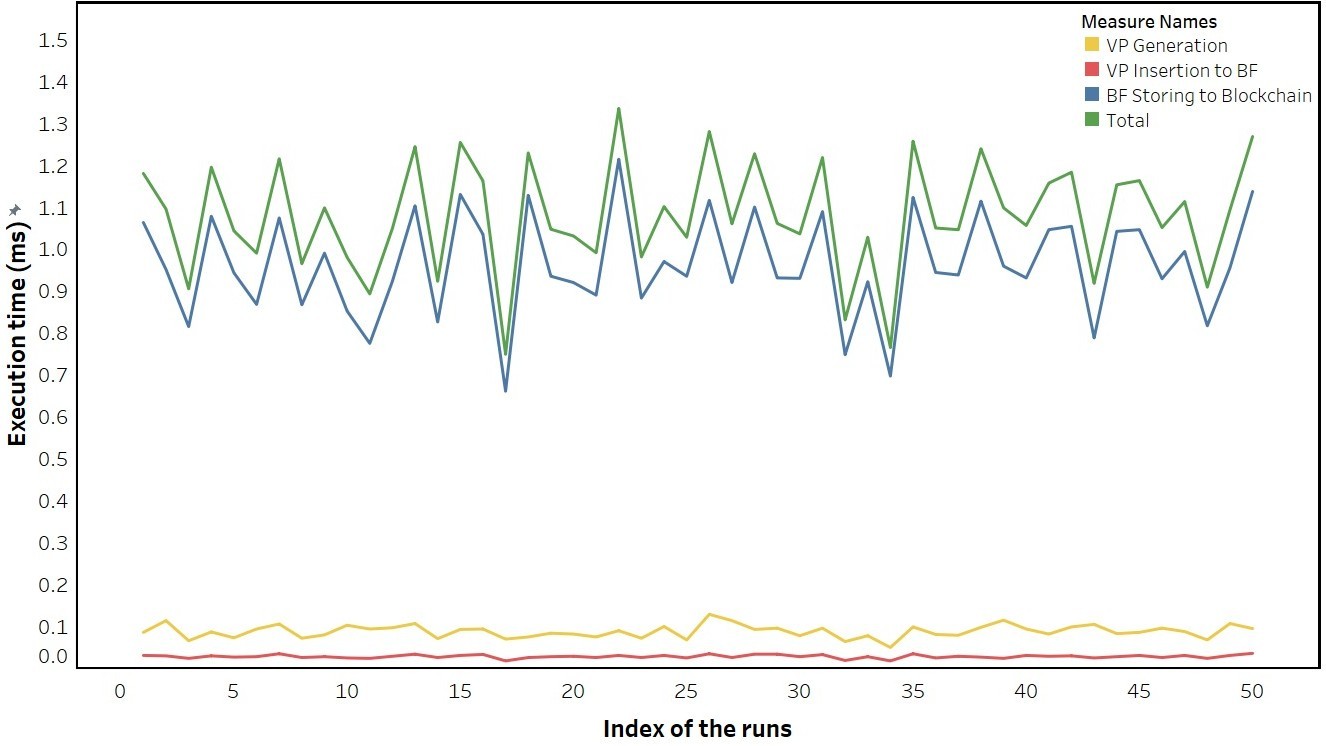}
         \caption{}
         \label{fig_vp_generation_pc}
     \end{subfigure}
      \begin{subfigure}{0.328\textwidth}
         \includegraphics[width=\linewidth]{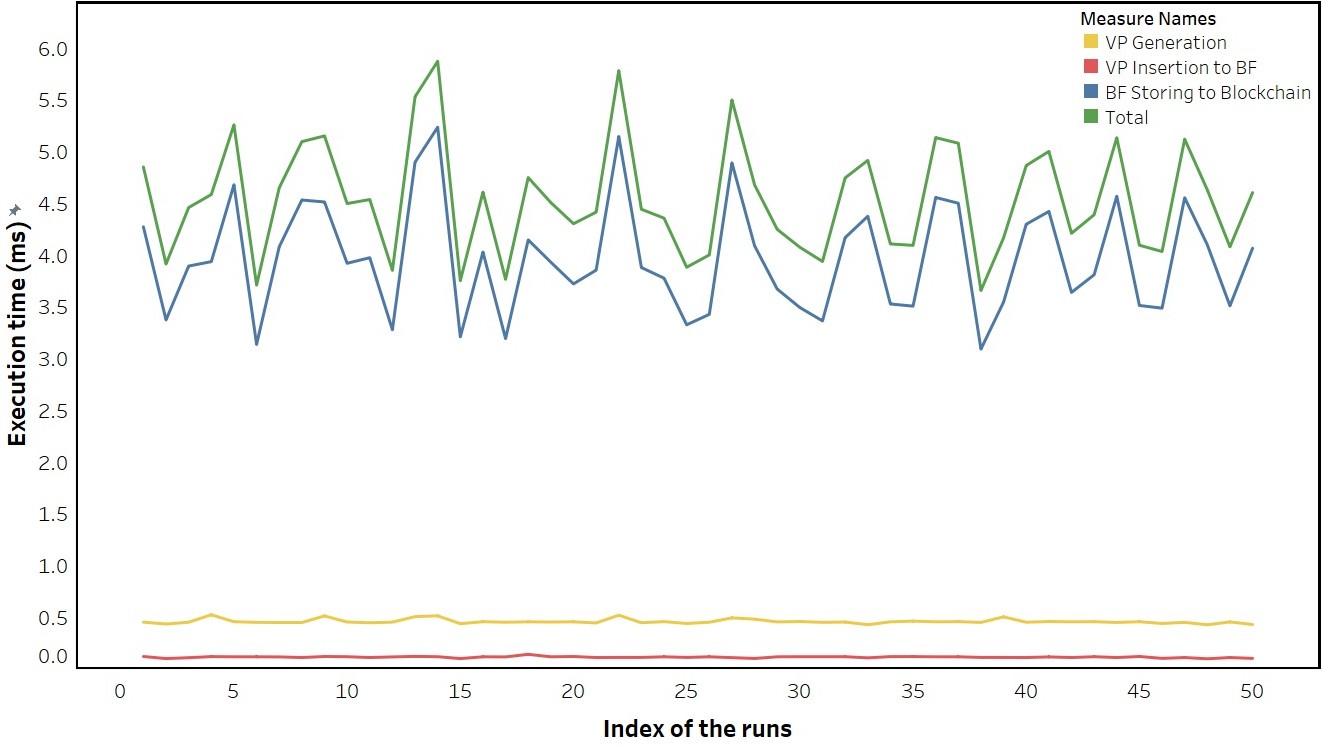}
         \caption{}
         \label{fig_vp_generation_rpi}
     \end{subfigure}
     \caption{VP generation time of (a) Gateway (b) PC and (c) RPI, for 50 runs}
     \label{fig_vp_generation_time}
     \vspace{-0.4cm}
\end{figure*}

Figures \ref{fig_vp_generation_time} (a), \ref{fig_vp_generation_time} (b) and \ref{fig_vp_generation_time} (c) illustrate 
the execution time of the $VP$ generation processes 
for the Gateway, PC and RPi devices, respectively, obtained from 50 runs.
In the figures, \emph{VP Computation} represents the time required to generate the new $GA_{i}^{j}$ and $VP_{i}^{j}$. \emph{VP Insertion to BF} represents the time required to insert the new $VP_{i}^{j}$ into $BF$. \emph{BF Storing to Blockchain} represents the time required to store $BF$ in the blockchain transaction. \emph{VP Generation} represents the overall time required to complete the VP generation processes. Table \ref{tab:generation} also summarizes the runtime of $VP$ generation processes for the three types of devices mentioned above.

\begin{table*}[thb]
    \centering
    \caption{Average VP generation time (in ms) for Gateway, PC and RPi}
    \begin{tabular}{|ccccc|}
        \hline
        \multirow{2}{*}{Devices} & \multicolumn{4}{|c|}{Operations}                                                                                        \\ \cline{2-5}
         & \multicolumn{1}{|c|}{VP computation} & \multicolumn{1}{c|}{VP insertion to BF} & \multicolumn{1}{c|}{BF storing to blockchain} & \textbf{Total} \\ \hline
         \multicolumn{1}{|c|}{Gateway} & \multicolumn{1}{|c|}{0.0596}         & \multicolumn{1}{c|}{0.0238}             & \multicolumn{1}{c|}{0.4509}                   & 0.5343         \\ \hline
          \multicolumn{1}{|c|}{PC} & \multicolumn{1}{|c|}{0.08852}         & \multicolumn{1}{c|}{0.02837}             & \multicolumn{1}{c|}{0.962}                   & 1.07887         \\ \hline
          \multicolumn{1}{|c|}{RPi} & \multicolumn{1}{|c|}{0.46152}         & \multicolumn{1}{c|}{0.11673}             & \multicolumn{1}{c|}{3.96281}                   & 4.54106         \\ \hline
    \end{tabular}
    \label{tab:generation}
\end{table*}

\begin{figure*}[htb]
     \centering
     \begin{subfigure}{0.328\textwidth}
         \centering
         \includegraphics[width=\linewidth]{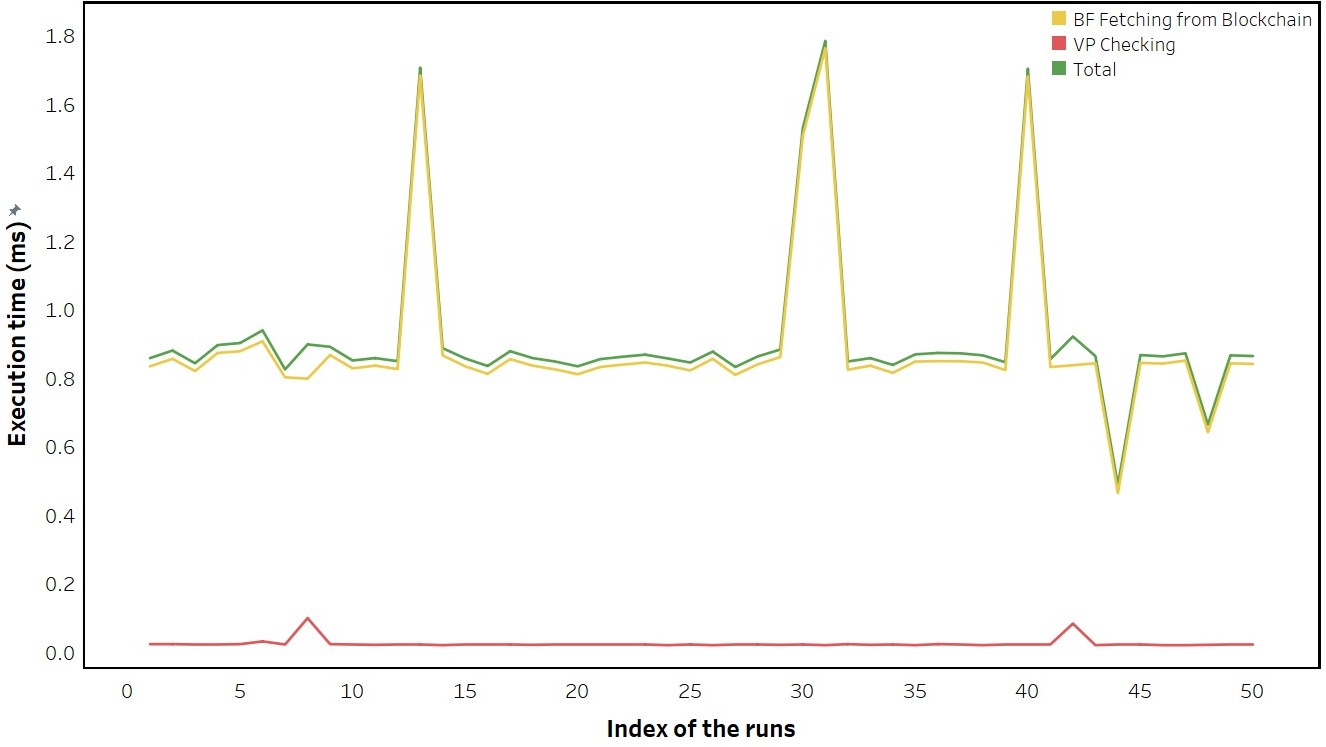}
         \caption{}
         \label{fig_vp_verification_gateway}
     \end{subfigure}
     \begin{subfigure}{0.328\textwidth}
         \centering
         \includegraphics[width=\linewidth]{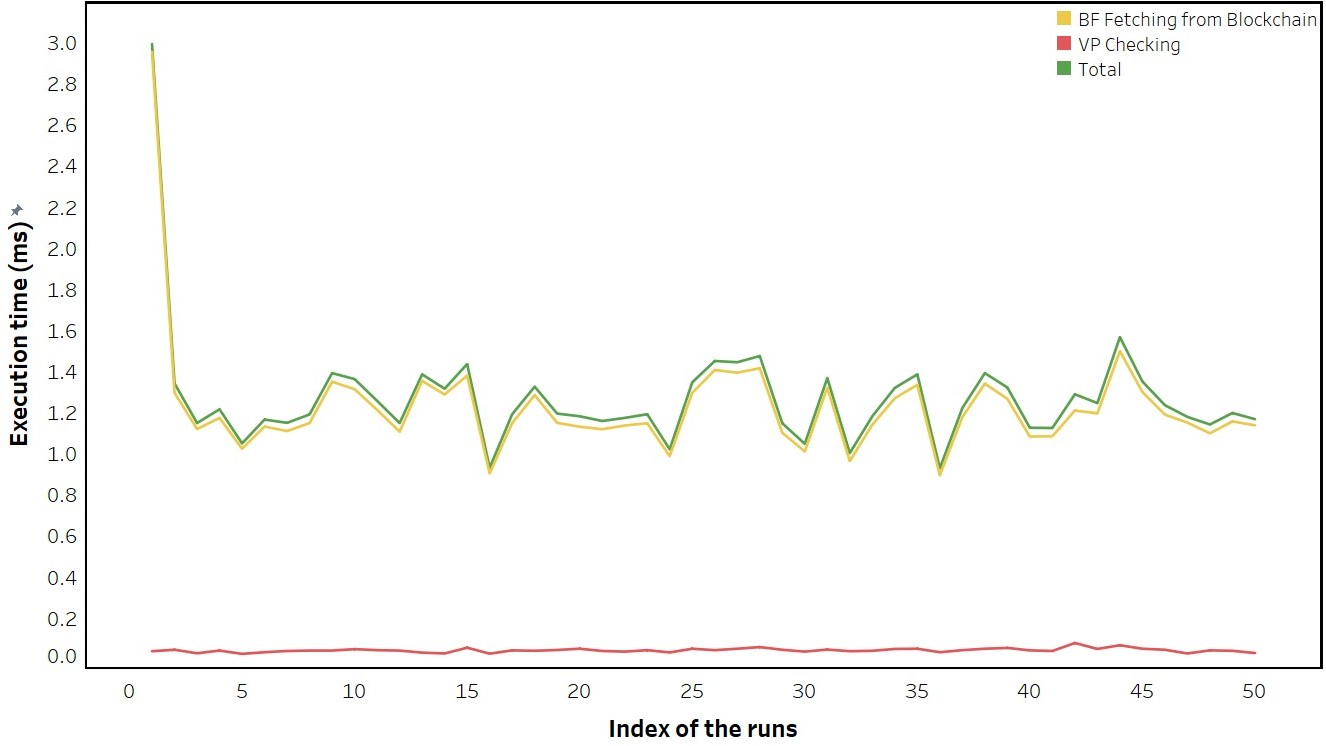}
         \caption{}
         \label{fig_vp_verification_pc}
     \end{subfigure}
      \begin{subfigure}{0.328\textwidth}
         \centering
         \includegraphics[width=\linewidth]{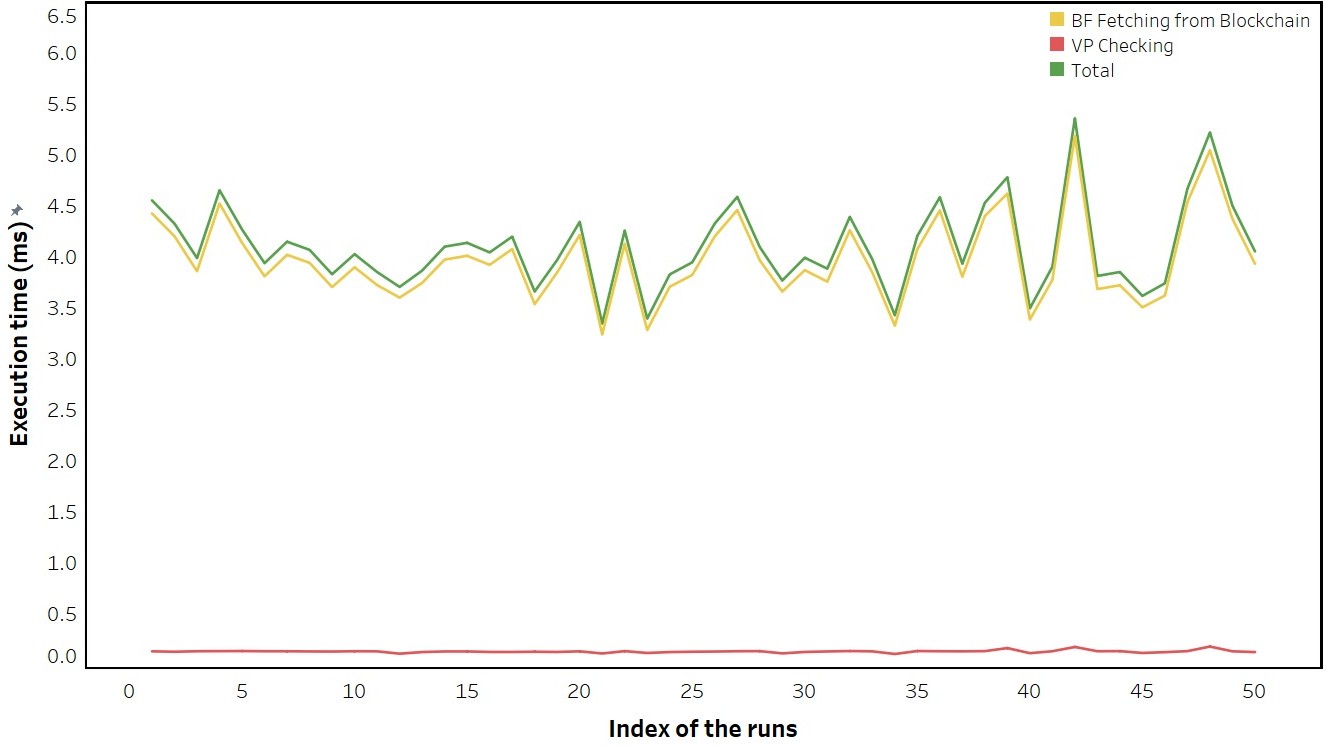}
         \caption{}
         \label{fig_vp_verification_rpi}
     \end{subfigure}
     \caption{VP verification time of (a) Gateway (b) PC and (c) RPI, for 50 runs}
     \label{fig_vp_verification_time}
\end{figure*}

Figures \ref{fig_vp_verification_time} (a),  \ref{fig_vp_verification_time} (b) and \ref{fig_vp_verification_time} (c) illustrate 
the execution time of the $VP$ verification processes 
for the Gateway, PC and RPi devices, respectively, obtained from 50 runs. 
From the figures, \emph{VP Fetching from Blockchain} represents the time required to fetch the latest $BF$ from blockchain. \emph{VP Checking} represents the verification time required to verify the $SGA_{i}$ with the latest $BF$, then returns access granted/denied based on the verification result. The \emph{VP Verification} represents the overall time required to complete the VP verification processes. 
Table \ref{tab:verification} summarizes the runtime of the $VP$ verification processes for the Gateway, PC and RPi devices.

The execution times for the processes outlined in Figures 
\ref{fig_vp_generation_time} (a) -- (c) and Figures \ref{fig_vp_verification_time} (a) -- (c) exhibit some fluctuation. This variability can be attributed to commonly existing conditions such as system loads, inefficient system scheduling policies, cache misses, and JVM garbage collection, as discussed in \cite{wang2012average}. However, based on the experimental results, while fluctuations do occur, they are infrequent.

On the other hand, the $VP$ checking time using the $BF$ remains constant regardless of the number of $VP$s stored. It is reflected in our experiment above wherein across 50 trials, the $VP$ verification time exhibited no constant increase. This observation confirms that the incorporation of additional $VP$ within the $BF$ does not impact the $VP$ verification time.

\begin{table}[thb]
    \centering
    \caption{Average VP verification time (in ms) for Gateway, PC and RPi}
    \begin{tabular}{|cccc|}
        \hline
        \multirow{2}{*}{Devices} & \multicolumn{3}{|c|}{ Operations} \\ \cline{2-4}
         & \multicolumn{1}{|c|}{BF fetching from blockchain} & \multicolumn{1}{c|}{VP checking} & \textbf{Total} \\ \hline
         \multicolumn{1}{|c|}{Gateway} & \multicolumn{1}{|c|}{0.8943} & \multicolumn{1}{c|}{0.0256} & 0.9199 \\ \hline
          \multicolumn{1}{|c|}{PC} & \multicolumn{1}{|c|}{1.22844} & \multicolumn{1}{c|}{0.04342} & 1.27186 \\ \hline
          \multicolumn{1}{|c|}{RPi} & \multicolumn{1}{|c|}{3.97237}  & \multicolumn{1}{c|}{0.12676} & 4.09913  \\ \hline
    \end{tabular}
    \label{tab:verification}
\end{table}

\subsection{Storage overhead}

We also assessed the storage overhead 
our scheme based on the experimental setup detailed in Section \ref{experimental_setup}. The primary storage cost in the proposed scheme arises from the utilization of the $BF$ as the core data structure. Other values, such as $VP$ and $GA$, are only temporarily utilized during generation. Once generated, all values except for the $BF$ are discarded from memory.

In our $BF$ initialization, the dynamically allocated bit array occupies 1199 bytes, while the $BF$ data structure itself uses 56 bytes. Given that the $BF$ is employed to store $VP$, a comparison without using the $BF$ reveals that each $VP$ consumes 32 bytes, as the MIRACL SHA-256 returns a 32-byte hashed value regardless of the input size. With our $BF$ initialization, capable of accommodating up to 1000 entries of $VP$, it effectively compresses 32,000 bytes of data into just 1255 bytes. This demonstrates the significant space efficiency achieved through the use of the $BF$ data structure.

\section{Conclusion}\label{conclusion}
Unauthorized access has been a critical security concern in today's digital world. The emerging of advanced exploitation techniques, such as session hijacking attacks, makes it even more challenging to alleviate the concern. Unfortunately, existing access control and authorization mechanisms are not sufficient to address such critical security concerns. 

In this paper, we proposed a multi-factor authorization (MFAz) scheme that provides a proactive mitigation strategy against a wide-range of unauthorized access attacks. The scheme uses access control rules (ARs) and systematically generated verification points (VPs) as first and second authorization factors, respectively, to raise the security bar against conventional and advanced (e.g., session hijacking) unauthorized access attacks. 

We implemented the scheme using cryptographic primitives, bloom filter (to achieve efficiency) and blockchain (to enhance availability and decentralization). Then, we tested and evaluated the scheme in a smart-city based testbed incorporating different devices with varying computational capacities. The evaluation results indicate that our scheme effectively mitigates conventional and session hijacking attacks and enhances the overall system security. Its efficiency (both runtime and storage overhead) is also experimentally validated and 
it is practical even in resource-constrained devices. 
Future research could explore further enhancements to our approach by integrating additional security factors and adapting them to address different threat vectors in other domains. 

\section*{Acknowledgment}

This research is supported in part by the National Research Foundation, Singapore, and the Ministry of National Development, Singapore under its Cities of Tomorrow R\&D Programme (CoT Award COT-V2-2021-1). Any opinions, findings and conclusions or recommendations expressed in this material are those of the author(s) and do not reflect the views of the National Research Foundation Singapore, and Ministry of National Development, Singapore.

\bibliographystyle{splncs04}
\bibliography{10_references}

\begin{thebibliography}{10}
\providecommand{\url}[1]{\texttt{#1}}
\providecommand{\urlprefix}{URL }
\providecommand{\doi}[1]{https://doi.org/#1}

\bibitem{ganache}
Ganache, \url{https://archive.trufflesuite.com/ganache/}

\bibitem{miracl}
Miracl (2018), \url{https://github.com/miracl}

\bibitem{addobea2023secure}
Addobea, A.A., Li, Q., Obiri~Jr, I.A., Hou, J.: Secure multi-factor access
  control mechanism for pairing blockchains. Journal of Information Security
  and Applications  \textbf{74},  103477 (2023)

\bibitem{almushiti2023investigation}
Almushiti, E., Zaki, R., Thamer, N., Alshaya, R.: An investigation of broken
  access control types, vulnerabilities, protection, and security. In:
  International Conference on Innovation of Emerging Information and
  Communication Technology. pp. 253--269. Springer (2023)

\bibitem{alsahlani2021lmaas}
Alsahlani, A.Y.F., Popa, A.: Lmaas-iot: Lightweight multi-factor authentication
  and authorization scheme for real-time data access in iot cloud-based
  environment. Journal of Network and Computer Applications  \textbf{192},
  103177 (2021)

\bibitem{angcsur2025}
Ang, K.W., Chekole, E.G., Zhou, J.: Unveiling the covert vulnerabilities in
  multi-factor authentication protocols: A systematic review and security
  analysis. ACM Comput. Surv.  \textbf{57}(11) (Jun 2025).
  \doi{10.1145/3734864}, \url{https://doi.org/10.1145/3734864}

\bibitem{armando2014selective}
Armando, A., Carbone, R., Chekole, E.G., Petrazzuolo, C., Ranalli, A., Ranise,
  S.: Selective release of smart metering data in multi-domain smart grids. In:
  Cuellar, J. (ed.) Smart Grid Security. pp. 48--62. Springer International
  Publishing, Cham (2014)

\bibitem{armando2014attribute}
Armando, A., Carbone, R., Chekole, E.G., Ranise, S.: Attribute based access
  control for apis in spring security. In: Proceedings of the 19th ACM
  symposium on Access control models and technologies. pp. 85--88 (2014)

\bibitem{banyal2013multi}
Banyal, R.K., Jain, P., Jain, V.K.: Multi-factor authentication framework for
  cloud computing. In: 2013 fifth international conference on computational
  intelligence, modelling and simulation. pp. 105--110. IEEE (2013)

\bibitem{bloom1970space}
Bloom, B.H.: Space/time trade-offs in hash coding with allowable errors.
  Communications of the ACM  \textbf{13}(7),  422--426 (1970)

\bibitem{broder2004network}
Broder, A., Mitzenmacher, M.: Network applications of bloom filters: A survey.
  Internet mathematics  \textbf{1}(4),  485--509 (2004)

\bibitem{buterin2014next}
Buterin, V., et~al.: A next-generation smart contract and decentralized
  application platform. white paper  \textbf{3}(37), ~2--1 (2014)

\bibitem{cahn2016empirical}
Cahn, A., Alfeld, S., Barford, P., Muthukrishnan, S.: An empirical study of web
  cookies. In: Proceedings of the 25th international conference on world wide
  web. pp. 891--901 (2016)

\bibitem{mfaa}
Chekole, E.G., Zhou, J.: Mfaa: Historical hash based multi-factor
  authentication and authorization in iiot. In: 2024 Annual Computer Security
  Applications Conference Workshops (ACSAC Workshops). pp. 133--144 (2024).
  \doi{10.1109/ACSACW65225.2024.00021}

\bibitem{dacosta2012one}
Dacosta, I., Chakradeo, S., Ahamad, M., Traynor, P.: One-time cookies:
  Preventing session hijacking attacks with stateless authentication tokens.
  ACM Transactions on Internet Technology (TOIT)  \textbf{12}(1),  1--24 (2012)

\bibitem{dowling2021cryptographic}
Dowling, B., Fischlin, M., G{\"u}nther, F., Stebila, D.: A cryptographic
  analysis of the tls 1.3 handshake protocol. Journal of Cryptology
  \textbf{34}(4), ~37 (2021)

\bibitem{hu2015attribute}
Hu, V.C., Kuhn, D.R., Ferraiolo, D.F., Voas, J.: Attribute-based access
  control. Computer  \textbf{48}(2),  85--88 (2015)

\bibitem{hwang2022web}
Hwang, W.S., Shon, J.G., Park, J.S.: Web session hijacking defense technique
  using user information. Human-centric Computing and Information Sciences
  \textbf{12}, ~16 (2022)

\bibitem{kaiser2022multi}
Kaiser, T., Siddiqua, R., Hasan, M.M.U.: A multi-layer security system for data
  access control, authentication, and authorization. Ph.D. thesis, Brac
  University (2022)

\bibitem{kamal2016state}
Kamal, P.: State of the art survey on session hijacking. Global Journal of
  Computer Science and Technology  \textbf{16}(1),  39--49 (2016)

\bibitem{xss}
Kaur, J., Garg, U., Bathla, G.: Detection of cross-site scripting (xss) attacks
  using machine learning techniques: a review. Artificial Intelligence Review
  \textbf{56}(11),  12725--12769 (2023)

\bibitem{khan2021blockchain}
Khan, S.N., Loukil, F., Ghedira-Guegan, C., Benkhelifa, E., Bani-Hani, A.:
  Blockchain smart contracts: Applications, challenges, and future trends.
  Peer-to-peer Networking and Applications  \textbf{14},  2901--2925 (2021)

\bibitem{kristol2001http}
Kristol, D.M.: Http cookies: Standards, privacy, and politics. ACM Transactions
  on Internet Technology (TOIT)  \textbf{1}(2),  151--198 (2001)

\bibitem{kwon2019security}
Kwon, H., Nam, H., Lee, S., Hahn, C., Hur, J.: (in-) security of cookies in
  https: Cookie theft by removing cookie flags. IEEE Transactions on
  Information Forensics and Security  \textbf{15},  1204--1215 (2019)

\bibitem{luo2018optimizing}
Luo, L., Guo, D., Ma, R.T., Rottenstreich, O., Luo, X.: Optimizing bloom
  filter: Challenges, solutions, and comparisons. IEEE Communications Surveys
  \& Tutorials  \textbf{21}(2),  1912--1949 (2018)

\bibitem{manjula2021pre}
Manjula, B.V.B., Naik, R.L.: Pre-authorization and post-authorization
  techniques for detecting and preventing the session hijacking. International
  Journal of Future Generation Communication and Networking  \textbf{14}(1),
  359--371 (2021)

\bibitem{muzammil2024unveiling}
Muzammil, M.B., Bilal, M., Ajmal, S., Shongwe, S.C., Ghadi, Y.Y.: Unveiling
  vulnerabilities of web attacks considering man in the middle attack and
  session hijacking. IEEE Access  (2024)

\bibitem{sidejacking}
Muzammil, M.B., Bilal, M., Ajmal, S., Shongwe, S.C., Ghadi, Y.Y.: Unveiling
  vulnerabilities of web attacks considering man in the middle attack and
  session hijacking. IEEE Access  \textbf{12},  6365--6375 (2024)

\bibitem{nakamoto2008bitcoin}
Nakamoto, S.: Bitcoin: A peer-to-peer electronic cash system  (2008)

\bibitem{ogundele2020detection}
Ogundele, I.O., Akinade, A.O., Alakiri, H.O., Aromolaran, A.A., Uzoma, B.:
  Detection and prevention of session hijacking in web application management.
  Int J Adv Res Comput Commun Eng  \textbf{9}(6),  1--10 (2020)

\bibitem{ometov2018multi}
Ometov, A., Bezzateev, S., M{\"a}kitalo, N., Andreev, S., Mikkonen, T.,
  Koucheryavy, Y.: Multi-factor authentication: A survey. Cryptography
  \textbf{2}(1), ~1 (2018)

\bibitem{brute_forcing}
Ruambo, F.A., Masanga, E.E., Lufyagila, B., Ateya, A.A., Abd El-Latif, A.A.,
  Almousa, M., Abd-El-Atty, B.: Brute-force attack mitigation on remote access
  services via software-defined perimeter. Scientific Reports  \textbf{15}(1),
  1--35 (2025)

\bibitem{samarati2000access}
Samarati, P., De~Vimercati, S.C.: Access control: Policies, models, and
  mechanisms. In: International school on foundations of security analysis and
  design, pp. 137--196. Springer (2000)

\bibitem{sandhu1998role}
Sandhu, R.S.: Role-based access control. In: Advances in computers, vol.~46,
  pp. 237--286. Elsevier (1998)

\bibitem{sathiyaseelan2017proposed}
Sathiyaseelan, A.M., Joseph, V., Srinivasaraghavan, A.: A proposed system for
  preventing session hijacking with modified one-time cookies. In: 2017
  International Conference on Big Data Analytics and Computational Intelligence
  (ICBDAC). pp. 451--454. IEEE (2017)

\bibitem{sauber2017session}
Sauber, A.J.: Session Armor: Protection Against Session Hijacking using
  Per-Request Authentication. Drexel University (2017)

\bibitem{singh2022handshake}
Singh, A.P., Singh, M.: Handshake comparison between tls v 1.2 and tls v 1.3
  protocol. In: Cyber Security in Intelligent Computing and Communications, pp.
  143--155. Springer (2022)

\bibitem{sivakorn2016cracked}
Sivakorn, S., Polakis, I., Keromytis, A.D.: The cracked cookie jar: Http cookie
  hijacking and the exposure of private information. In: 2016 IEEE Symposium on
  Security and Privacy (SP). pp. 724--742. IEEE (2016)

\bibitem{sloan2017unauthorized}
Sloan, R., Warner, R.: Unauthorized access: The crisis in online privacy and
  security. Taylor \& Francis (2017)

\bibitem{session_fixation}
Squarcina, M., Ad{\~a}o, P., Veronese, L., Maffei, M.: Cookie crumbles:
  breaking and fixing web session integrity. In: 32nd USENIX Security Symposium
  (USENIX Security 23). pp. 5539--5556 (2023)

\bibitem{squarcina2023cookie}
Squarcina, M., Ad{\~a}o, P., Veronese, L., Maffei, M.: Cookie crumbles:
  breaking and fixing web session integrity. In: 32nd USENIX Security Symposium
  (USENIX Security 23). pp. 5539--5556 (2023)

\bibitem{wang2012average}
Wang, Q., Kanemasa, Y., Kawaba, M., Pu, C.: When average is not average: large
  response time fluctuations in n-tier systems. In: Proceedings of the 9th
  international conference on Autonomic computing. pp. 33--42 (2012)

\bibitem{wang2019blockchain}
Wang, S., Ouyang, L., Yuan, Y., Ni, X., Han, X., Wang, F.Y.: Blockchain-enabled
  smart contracts: architecture, applications, and future trends. IEEE
  Transactions on Systems, Man, and Cybernetics: Systems  \textbf{49}(11),
  2266--2277 (2019)

\bibitem{wee2024excavating}
Wee, A.K., Chekole, E.G., Zhou, J.: Excavating vulnerabilities lurking in
  multi-factor authentication protocols: A systematic security analysis. arXiv
  preprint arXiv:2407.20459  (2024)

\bibitem{yao2020dynamic}
Yao, Q., Wang, Q., Zhang, X., Fei, J.: Dynamic access control and authorization
  system based on zero-trust architecture. In: Proceedings of the 2020 1st
  international conference on control, robotics and intelligent system. pp.
  123--127 (2020)

\bibitem{zong2019policy}
Zong, Y., Guo, Y., Chen, X.: Policy-based access control for robotic
  applications. In: 2019 IEEE International Conference on Service-Oriented
  System Engineering (SOSE). pp. 368--3685. IEEE (2019)

\end{thebibliography}


\end{document}